%% file: Incidence.tex
\documentclass[12pt,a4paper]{article}

\newif\ifpdf
\ifx\pdfoutput\undefined
    \pdffalse                   
\else
    \ifx\pdfoutput\relax
        \pdffalse               
    \else
        \ifnum\pdfoutput>0
            \pdftrue            
        \else
            \pdffalse           
        \fi
    \fi
\fi

\ifpdf
    \usepackage[pdftex]{graphicx}
    \DeclareGraphicsExtensions{.pdf, .jpg, .tif}
\else
    \usepackage{graphicx}
    \DeclareGraphicsExtensions{.eps, .jpg}
\fi

\usepackage{amsfonts}
\usepackage{amssymb}
\usepackage{amsthm}
\usepackage{mathrsfs}
\usepackage{enumerate}
\usepackage{amsmath}
\usepackage{pgf}

\begin{document}

\bibliographystyle{unsrt}

\title{Relating Recent Infection Prevalence to
Incidence with a Sub-population of Non-progressors.}

\author{Thomas A. McWalter\footnote{School of Computational and
Applied Mathematics, University of the Witwatersrand, South
Africa} $ $ and Alex Welte\footnote{Corresponding author; DST/NRF
Centre of Excellence in Epidemiological Modelling and Analysis
(SACEMA) and School of Computational and Applied Mathematics,
University of the Witwatersrand, South Africa}}

\maketitle

\input{abstract}

\vspace{7cm}

\pagebreak

\input{body}

\section*{Acknowledgements}

The authors wish to thank Robert de Mello Koch, Paul Fatti, John
Hargrove, Norman Ives and Brian Williams for useful discussions.

\input{appendix}

\bibliography{Incidence}

\input{figures}

\end{document}

%% file: abstract.tex
\abstract{We present a new analysis of relationships between disease incidence and the prevalence of an experimentally defined state of `recent infection'. This leads to a clean separation between \emph{biological parameters} (properties of disease progression as reflected in a test for recent infection), which need to be calibrated, and \emph{epidemiological state variables}, which are estimated in a cross-sectional survey. The framework takes into account the possibility that details of the assay and host/pathogen chemistry leave a (knowable) fraction of the population in the recent category for all times. This systematically addresses an issue which is the source of some controversy about the appropriate use of the BED assay for defining recent HIV infection. Analysis of relative contributions of error arising variously from statistical considerations and simplifications of general expressions indicate that statistical error dominates heavily over all sources of bias for realistic epidemiological and biological scenarios. Numerical calculations validate the approximations made in analytical relations.}

%% file: body.tex

\newcommand{\Hsub}{\textrm{H}}
\newcommand{\Usub}{\textrm{U}}
\newcommand{\Osub}{\textrm{O}}
\newcommand{\Asub}{\textrm{A}}
\newcommand{\Rsub}{\textrm{R}}
\newcommand{\Lsub}{\textrm{L}}
\newcommand{\PUsub}{\textrm{PU}}

\newcommand{\fua}{f_{\scriptscriptstyle \Usub|\Asub}}
\newcommand{\Fua}{F_{\scriptscriptstyle \Usub|\Asub}}
\newcommand{\Sua}{S_{\scriptscriptstyle \Usub|\Asub}}
\newcommand{\tauua}{\tau_{\scriptscriptstyle \Usub|\Asub}}

\newcommand{\fa}{f_{\scriptscriptstyle \Asub}}
\newcommand{\Fa}{F_{\scriptscriptstyle \Asub}}
\newcommand{\Sa}{S_{\scriptscriptstyle \Asub}}
\newcommand{\taua}{\tau_{\scriptscriptstyle \Asub}}

\newcommand{\fu}{f_{\scriptscriptstyle \Usub}}
\newcommand{\Su}{S_{\scriptscriptstyle \Usub}}
\newcommand{\tauu}{\tau_{\scriptscriptstyle \Usub}}

\newcommand{\fpua}{f_{\scriptscriptstyle \PUsub|\Asub}}
\newcommand{\Spua}{S_{\scriptscriptstyle \PUsub|\Asub}}

\newcommand{\fra}{f_{\scriptscriptstyle \Rsub|\Asub}}
\newcommand{\Fra}{F_{\scriptscriptstyle \Rsub|\Asub}}
\newcommand{\Sra}{S_{\scriptscriptstyle \Rsub|\Asub}}

\newcommand{\Sr}{S_{\scriptscriptstyle \Rsub}}
\newcommand{\taur}{\tau_{\scriptscriptstyle \Rsub}}

\newcommand{\Iw}{I_{\scriptscriptstyle \textrm{W}}}
\newcommand{\Iest}{I_{\scriptscriptstyle \textrm{est}}}


\renewcommand{\H}{H}
\newcommand{\U}{U}
\renewcommand{\O}{O}
\newcommand{\R}{R}


\newcommand{\tPU}{\textrm{P}_{\scriptscriptstyle \Usub}}
\newcommand{\tPO}{\textrm{P}_{\scriptscriptstyle \Osub}}
\newcommand{\tNP}{\textrm{NP}}
\newcommand{\tNPR}{{\tNP}_{\scriptscriptstyle \Rsub}}
\newcommand{\tNPL}{{\tNP}_{\scriptscriptstyle \Lsub}}


\newcommand{\PU}{P_{\scriptscriptstyle \Usub}}
\newcommand{\PO}{P_{\scriptscriptstyle \Osub}}
\newcommand{\NP}{\textit{NP}}
\newcommand{\NPR}{{\NP}_{\scriptscriptstyle \Rsub}}
\newcommand{\NPL}{{\NP}_{\scriptscriptstyle \Lsub}}


\renewcommand{\P}{\mathbb{P}}
\newcommand{\PrH}{\P_{{\scriptscriptstyle \Hsub}}}
\newcommand{\PrU}{\P_{{\scriptscriptstyle \Usub}}}
\newcommand{\PrO}{\P_{{\scriptscriptstyle \Osub}}}
\newcommand{\PrX}{\P_{{\scriptscriptstyle \textrm{X}}}}
\newcommand{\NH}{N_{\scriptscriptstyle \Hsub}}
\newcommand{\NU}{N_{\scriptscriptstyle \Usub}}
\newcommand{\NO}{N_{\scriptscriptstyle \Osub}}
\newcommand{\NX}{N_{\scriptscriptstyle \textrm{X}}}

\newcommand{\Pnp}{\P_{{\scriptscriptstyle \textrm{NP}}}}

\newcommand{\E}[1]{\mathbb{E}\left[#1\right]}       

\newcommand{\figref}[1]{Figure~\ref{#1}}

\section{Introduction}

Reliable estimation of disease \emph{incidence} (rate of
occurrence of new infections) and \emph{prevalence} (the fraction
of a population in an inflected state) are
central to the determination of epidemiological trends, especially
for the allocation of resources and evaluation of interventions.
Prevalence estimation is relatively straightforward, for example
by cross-sectional survey. Incidence estimates are notoriously
problematic, though potentially of crucial importance. An
approximate measure of incidence in a population is required for
the proper planning of sample sizes and costing for clinical
trials and other population based studies. Repeated follow-up of a
representative cohort is often touted as the `gold standard' for
estimating incidence, but is costly, time intensive and still
prone to some intrinsic problems. For example, there may be bias in the factors determining which subjects are
lost to follow-up. Furthermore, ethical considerations demand that
a cohort study involve substantial support for subjects to avoid
becoming infected, which may make the cohort unrepresentative of
the population of interest.

Numerous methods have been proposed for inferring incidence from
single or multiple cross-sectional surveys rather than following
up a cohort
\cite{Brookmeyer:1995:1,Posner:2002:1,cdcjama:1,bed:2002:1,parekh:2001:1,Brookmeyer:2006:1,
nascimento:2004:1,bed:2006:1,wong:2006:1,jhbed:1}. A central idea
in most of these \cite{Brookmeyer:1995:1,cdcjama:1,bed:2002:1,parekh:2001:1,Brookmeyer:2006:1,bed:2006:1,jhbed:1} is to count the prevalence of a state of
`recent infection', which naturally depends on the recent
incidence. The relationship between the two is in general not
simple and depends in detail on the recent population dynamics as
well as distributions which capture the inter-subject variability
of progression through stages of infection, as they are observed
by the specific laboratory assays used in the test for recent
infection (TRI). For this approach to be sensible, a working
definition of `recent infection' must be calibrated, for example
by repeatedly following up subjects over a period during which
they become infected. This is effectively as much effort as one
measurement of incidence by follow-up. The calibration is then used to infer incidence from each of many subsequent
independent cross-sectional surveys.

Owing to the devastating impact of the HIV epidemic, and the many
challenges of research and intervention design, the problem of
estimating HIV incidence has attracted considerable interest in
recent years.  The prospect of using a TRI is in principle very
attractive. Given the range of values of incidence likely to be
observed in populations with a major epidemic (say 1-10\% per annum) a
mean definition of `recent' of approximately half a year is
desirable to yield reasonable statistical confidence for sample
sizes of a few thousand. The BED assay is currently the leading
candidate for such a test, but controversy has arisen about the
possibility of conducting a reliable calibration. This stems from
the fact that a subset of individuals (approximately $5\%$
\cite{bed:2006:1,jhbed:1}, potentially variable between viral and host
populations) fail to progress above any statistically useful
threshold set on the assay in the definition of `recent'
infection. This subset of individuals, who consequently remain
classified as `recently infected' for all times, poses a problem to
which there is currently no consensus remedy.

We present a new analysis of the interaction between
epidemiological trends and a model of inter-subject variability of
progression through an experimental category of `recent
infection'. Our model yields simple formulae for inference
\emph{even when a fraction of the population fails to progress out
of the recent category}. The only physiological assumption
required to deal with the non-progressors is that their survival
after infection is the same as the progressors. This assumption is also implicit in previous work on using the BED assay to estimate incidence.


A key conceptual point about our analysis, which distinguishes it from all others of which we are aware, is that we confront the fundamental limitation of what can be inferred from a cross-sectional survey. In particular, even perfect knowledge of an instantaneous population state does not uniquely determine the instantaneous incidence. At best, a weighted average of recent incidence can be inferred. Although the discrepancies between this weighted average and instantaneous incidence can be shown to be small compared to statistical errors (for our application), it can in principle be systematically incorporated into estimation of trends from multiple cross-sectional surveys.

The article is organized as follows. In Section 2 we first develop
a basic continuous time model defined by a time dependent
incidence and susceptible population, a distribution of times
after infection spent under the threshold on a TRI and a distribution
of post-infection survival times. We note that there is in
principle no specific relationship between the instantaneous
incidence and the prevalence of individuals who are infected and under the threshold. At best, one obtains a relationship between the prevalence of
under-threshold individuals and a convolution of the recent
incidence with a specific weighting function which is implied by
the use of a TRI. This relationship in principle includes all
moments of the distribution of the waiting times that individuals
spend under the threshold. We show that, for realistic rates of
variation in the susceptible population, only the mean of the
waiting time distribution is needed, and a simple expression for a
weighted average of the incidence is obtained. The basic model is
extended to allow some fraction of individuals (specified by a new parameter)
to be assigned infinite waiting times under the threshold of the
TRI. This leads to only very minor modifications of the previous
expression for weighted incidence, namely a systematic
`subtraction' of over-counted `not recently infected' individuals
which are included in the experimental category `under threshold'.
This subtraction is similar, but not identical, to that proposed in \cite{jhbed:1}.

Section 3 explores the consequences of designing a cross-sectional
survey with a sample size $N$ based on the relations derived in
Section 2. Using a systematic expansion of the incidence estimator
in powers of $1/\sqrt{N}$ (derived in the Appendix), we note
consistency of the estimator (no bias in the limit of large $N$)
and derive an approximate expression for its relative variance.
These expressions facilitate error estimation both from a study
design and data analysis point of view. On calibration, we note
that trends, as opposed to absolute values, for incidence can be
obtained without any information about the distribution of finite
waiting times under the threshold. However, an estimate of the
fraction of non-progressors is essential. A key observation is
that, for realistic population dynamics and sample sizes, statistical error is much larger than bias.

Numerical simulations are presented in Section 4. These
demonstrate that the approximate statistical analysis of Section 2
is essentially correct, with discrepancies of the size expected
from the $1/\sqrt{N}$ expansion.

In the conclusion, we note that the framework presented here is quite general and is applicable to any TRI, as long as any finite probability for non-progression can be calibrated, survival is the same for progressors and non-progressors, and there is no `relapse' from over to under the recency threshold. It may be possible to modify the analysis to relax these requirements. We point out that the relationship between the present analysis and other proposals for using the BED TRI for measuring HIV incidence should be more systematically investigated. Some preliminary work in this direction has already been produced \cite{McWalterWelte2008B}.


\section{Relating the Prevalence of `Recent Infection' to Incidence}

We now outline a quite general approach to relating the key
demographic, epidemiological and biological processes which are
relevant to the estimation of incidence from cross-sectional
surveys of the prevalence of `recent infection'. This refines the
naive intuition that a high prevalence of `recently infected'
individuals means a high incidence.

\subsection*{The Basic Model}

A test for recent infection, such as the CDC STARHS algorithm, is typically obtained through the administration of two assays of different sensitivity. The more sensitive test distinguishes infected from healthy individuals and the less sensitive test, applied to the infected individuals, distinguishes `recent' from `long' established infection.

Consider an assay which yields a quantitative result, the value of which typically increases with time from infection. The BED assay is of this type. The quantitative result is a normalized optical density (ODn), which is an increasing function of the proportion of HIV-1 specific IgG. The hypothetical ongoing observation of an individual might lead to a curve similar to that displayed in \figref{ODcurvedist}A. Such an assay becomes the less sensitive component of a  test for recent infection when we declare a threshold value and define `observed to be recently infected' to be a test value under the threshold.

In practice, there is inevitably inter-individual variation in these progression curves. Plotting the curves for multiple individuals on a single graph would lead to something like \figref{ODcurvedist}B. Clearly, the category `observed to be recently infected' is not sharply defined by a time boundary, and we now adopt the more precise labels \emph{under} threshold (\U) and \emph{over} threshold (\O). The variability of times spent in the under-threshold category, conditional on being alive long enough to reach the threshold, is captured by a distribution of waiting times $\fua$ which may look something like that shown in \figref{ODcurvedist}C.

It is now possible to construct the basic epidemiological model shown in \figref{flowmodels}A. Since our analysis will focus on a variety of \emph{survival} functions $S(t)$, we shall refer to the susceptible population as the \emph{healthy} population $\H(t)$. Upon infection, individuals move from the healthy population to the under-threshold infected population $\U(t)$. Those that live long enough, reach the threshold after a waiting time distributed according to $\fua$, and enter the over-threshold population $\O(t)$. We denote by $\tauua$ a waiting time generated by the density $\fua$. The corresponding cumulative probability function is given by
\begin{equation}
    \Fua(t)=\int_0^t\fua(s)\,ds,
\end{equation}
while the probability of `survival' (persistence) in the
population \U, conditional on being alive, is
\begin{equation}
    \Sua(t)=\P(\tauua>t)= 1-\Fua(t),
\end{equation}
and the mean waiting time is
\begin{equation}
    \E{\tauua}=\int_0^\infty\tau\fua(\tau)\,d\tau=\int_0^\infty\Sua(t)\,dt.
\end{equation}
Analogously, we define $\fa$, $\taua$, $\Fa$, $\Sa$ and
$\E{\taua}$ in order to capture survival times (how long individuals remain \emph{alive}
after the moment of infection). We assume that survival time and
waiting time to threshold are independent in this model. Hence,
the probability, at a time delay $\Delta t$ after infection, of
being \emph{simultaneously} alive and under the threshold on the
assay is
\begin{align}
    \P(\taua>\Delta t\,\,\mbox{\tiny AND}\,\,\tauua>\Delta t)&=\Su(\Delta t)
    =\Sa(\Delta t)\Sua(\Delta t).
\intertext{Similarly, the probability of being
\emph{simultaneously} alive and over the threshold is}
    \P(\taua>\Delta t\,\,\mbox{\tiny AND}\,\,\tauua\leq\Delta t)&=
    \Sa(\Delta t)(1-\Sua(\Delta t)).
\end{align}
Hence, the mean time spent in the category {\U}, accounting for both progression and mortality, is $\E{\tauu}$.

New infections are generated by a non-homogeneous Poisson process with an intensity (probability per unit time of new arrivals) $\lambda(t)$. Let the instantaneous incidence be given by $I(t)$. Then, in a period $dt$ around time $t$, the expected number of new cases $dC$ is given by
\begin{equation}
    dC=\lambda(t)\,dt=I(t)\H(t)\,dt.\label{infectionslambda}
\end{equation}
We can now write down numerous expressions resulting from the model. For example, the expected number of historically accumulated cases up until time $t$ is given by
\begin{equation}
    C(t)=\int_{-\infty}^{t}\lambda(s)\,ds=\int_{-\infty}^{t}I(s)\H(s)\,ds.
\end{equation}
The expected populations of infected persons under and over
the threshold at time $t$ are
\begin{align}
    \U(t)&=\int_{-\infty}^{t}I(s)\H(s)\P(\taua>t-s\,\,\mbox{\tiny AND}\,\,\tauua>t-s)\,ds\notag\\
        &=\int_{-\infty}^{t}I(s)\H(s)\Su(t-s)\,ds\label{btag}
    \intertext{and}
    \O(t)&=\int_{-\infty}^{t}I(s)\H(s)\P(\taua>t-s\,\,\mbox{\tiny AND}\,\,\tauua\leq t-s)\,ds\notag\\
    &=\int_{-\infty}^{t}I(s)\H(s)\Sa(t-s)(1-\Sua(t-s))\,ds.\label{atag}
\end{align}

Our goal is to relate $I$ for recent times to instantaneous values
of $\H$, $\U$ and $\O$. We wish to emphasize the cautionary note that
there is \emph{fundamentally} a loss of information when one tries
to characterize  the history of a population based on observations
made at a single time point. The recent historical course of a
population, and even \emph{instantaneous} values of state
variables which are \emph{rates}, like incidence, are in general
not inferable from \emph{counting} data obtained in a single
survey. This is due to the fact that counts are, unavoidably,
convolutions of historical epidemiological variables, as in
\eqref{btag} and \eqref{atag} above. Any attempt to derive
incidence estimates from the counting of infections accumulated in
the recent past faces this problem, and at best some sort of
weighted average of the recent values of incidence can be inferred
without additional assumptions.

In general, a well defined construction of an estimate for incidence, based on data obtained in a survey conducted at time $t$, will be some sort of weighted average of past values
\begin{equation}
    \Iw(t)=\frac{\int_{-\infty}^t I(s)W(s,t)\,ds}{\int_{-\infty}^tW(s,t)\,ds},\label{iwdef}
\end{equation}
where $W(s,t)$ is a statistical weight arising from a convolution
of population history and biology. Since our goal is to estimate
incidence from a count of recently infected individuals, a natural
weighting function is one that reflects the relative contributions
to this count made by infections from different times in the
recent past. Hence, we consider
\begin{equation}
    W(s,t)=\H(s)\Su(t-s)\label{weight}
\end{equation}
since $W(s,t)$ is proportional to the probability that individuals are
\begin{enumerate}
    \item available for being infected at time $s<t$, and
    \item still alive and classified as under the threshold at time $t$, if infected at time $s$.
\end{enumerate}
Using \eqref{weight} as the weighting function leads to an
expression for the incidence given by
\begin{align}
    \Iw(t)&=\frac{\int_{-\infty}^{t}I(s)\H (s)\Su(t-s)\,ds}{\int_{-\infty}^{t}\H(s)\Su(t-s)\,ds}\notag\\
    &=\frac{\U(t)}{\int_{-\infty}^{t}\H(s)\Su(t-s)\,ds}\label{basic}.
\end{align}
The numerator in this expression is an instantaneous state
variable, while the denominator in principle involves data from
the entire history of the system as well as full knowledge of the
survival function $\Su$.

We will shortly show how to obtain a systematic approximation of the denominator, but a few remarks are in order about the practical meaning of this weighted average. In the case of constant incidence, the weighted average is the instantaneous value. In the case of a narrowly peaked distribution $\fua$, a constant rate of change of $I$ and a constant healthy population, the weighted average is approximately equal to the instantaneous incidence at a time $\E{\tauu}/2$ prior to the cross-sectional survey. If trends are fitted to the results of multiple cross-sectional surveys, this time lag could be more systematically accounted for.

\subsection*{A Simple Expression for Incidence}

A simplified expression for weighted incidence in terms of sample
and calibration data is now derived. We express the healthy
population using the expansion
\begin{equation}
    \H(t+s)=\H(t)+\H_1s+\H_2s^2+\ldots
\end{equation}
and use the identity
\begin{equation}
    \int_{-\infty}^0 s^n\Su(-s)\,ds=\tfrac{(-1)^n}{n+1}\E{\tauu^{n+1}},
\end{equation}
which follows directly from integration by parts. It then follows that the weighted incidence \eqref{basic} can be expressed as
\begin{align}
    \Iw(t)&=\frac{\U(t)}{\int_{-\infty}^{t}\H(s)\Su(t-s)\,ds}\label{isibdef}\\
    &=\frac{\U(t)}{\int_{-\infty}^0H(t+s)\Su(-s)\,ds}\notag\\
    &=\frac{\U(t)}{\H(t)\E{\tauu}-\tfrac{\H_1}{2}\E{\tauu^2}+\ldots}.\notag
\end{align}

If the healthy population is approximately constant for the
times where the weight $W$ is non-vanishing, we obtain the simple
relation
\begin{equation}
    \Iw=\frac{\U(t)}{\H(t)\,\E{\tauu}} \label{Isibsimple},
\end{equation}
which gives a weighted recent incidence in terms of instantaneous
state variables ($\H$ and $\U$) and the expected waiting time in
the under-threshold category.

Expectation values of the form $\E{\tauu^{n}}$ are not state
variables and should in principle be measured independently of a
particular cross-sectional survey. Usually this would be
accomplished in a calibration cohort follow-up study. Thus, after
calibrating some of these expectation values, we can deal with
a truncated expansion for $\H$ without further assumptions about
the behavior of $I$.

Some cautionary comments on calibration are, however, necessary. It seems unlikely that accurate measurements for terms of higher order than just $\E{\tauu}$ are practically possible for the case of the BED assay. Finding the non-leading terms $\H_i$ (for $i\geq1$) in the expansion of $\H$ will also require considerable demographic research.

These considerations mean that it is most likely that we will be
constrained to use the simple expression \eqref{Isibsimple} even
if the healthy population is not approximately constant over
the times where $W$ is non-vanishing. The key question then is:
how severe is the bias introduced by using the simple formula
under realistic non-constancy of the healthy population?

In order to explore this issue systematically we consider a non-constant healthy population given by $\H(t)=\H_0e^{\alpha t}$. This has a conveniently tunable degree of failure to conform with the constancy assumption required for \eqref{Isibsimple}. When $\alpha=0$ we have a constant number of healthy individuals, while a value of $\alpha=\ln(x)$ means the population grows by a factor of $x$ in one year. We can provide a survival function $\Su(t)$ for time measured in years, roughly inspired by the ODn progression on the BED assay, by specifying $\fu$ to be a Weibull distribution with scale parameter $l=0.57$ and shape parameter $k=8$. We now numerically evaluate the denominator of \eqref{isibdef} and compare it to the denominator of the simple formula \eqref{Isibsimple}. Note that this bias calculation is independent of the actual time dependance in $I$.

In \figref{bias}, the bias in the naive denominator (reported as a
fraction of the unbiased denominator) is shown as a function of
$\alpha$, reported as the annual percentage growth. Note that for
a population growing at $4\%$ per year, the bias is about 1.1\%.
As we shall see later when analyzing a slightly more complex model
of a TRI, this is small compared with the statistical error that
arises as a result of using realistic sample sizes. Thus, bias
arising from the non-constancy of the healthy population is
not a key concern unless there is very dramatic variation in
$\H$. The bias calculation demonstrated here is also applicable
to the more complex model that now follows.

\subsection*{Modeling Non-progressing Individuals}

A complicating factor for the BED assay is the fact that a small
number of individuals utterly fail to progress beyond any
practical ODn threshold used to define recency. This is due to
individual variation in biochemical details such as immune
response, for example. The non-progression phenomenon leads to a
long term accumulation of \emph{apparently} recently infected
individuals, as classified by the TRI, even though many of them
have been infected for a long time. There is currently no
consensus on how to handle this complication. We now generalize
the previous analysis to the situation where some individuals fail
to progress to the over-threshold category.

Consider the simple model captured in \figref{flowmodels}B. At the moment of infection, individuals transition from the healthy population to either a non-progressing population ($\tNP$) or to a progressing under-threshold population ($\tPU$). The probability of non-progression is $\Pnp$, and hence the probability of progression is $1-\Pnp$. Those individuals in $\tPU$ wait for a stochastic delay after which they move into the progressing over-threshold category $\tPO$. In the previous model, $\fua$ was the distribution of waiting times governing the transition, but since the waiting times for non-progressing individuals are infinite, $\fua$ cannot be normalized. Therefore, in order to specify the transition times from $\tPU$ to $\tPO$ in terms of a normalized density, we introduce the density of waiting times in the state of being a progressor and \emph{under} the threshold, conditional on being a \emph{alive}, denoted by $\fpua$. Then $\Sua(t)$, $\Spua(t)$ and $\Pnp$ are related by
\begin{equation}
    \Sua(t)=(1-\Pnp)\Spua(t)+\Pnp.
\end{equation}

The difficulty is that the TRI will classify as `recently infected' all the individuals in the $\tNP$ \emph{and} $\tPU$ categories even though some potentially large number in $\tNP$ are long infected. This can systematically be addressed by the following two key steps.

Firstly, we assume that the same survival function $\Sa$ is applicable to both progressing and non-progressing individuals. This is true if the differences between individuals which account for progression versus non-progression do not translate into significant differences in post-infection survival. This assumption has also been made, at least implicitly, in previous work on use of the BED assay for estimating HIV incidence (for example, see \cite{McWalterWelte2008B} for analysis of \cite{bed:2006:1}). Its applicability should in principle be tested, but we are unaware of any evidence suggesting that it is false.

Secondly, we introduce two artificial categories by separating the non-progressing population into `recently infected' ($\tNPR$) and `long infected' ($\tNPL$) sub-populations. Individuals entering the $\tNPR$ sub-population are assigned a waiting time drawn from $\fpua$ after which they transition to the $\tNPL$ category. Note that this is a book-keeping device used for convenience and, unlike the assumption about survival, does not rely on any property of disease progression. It is now possible to provide a sensible definition for the class of `recently infected individuals' (\R) which has a population given by
\begin{equation}
    \R(t)=\PU(t)+\NPR(t).\label{Req}
\end{equation}
Note that, since both $\tPU$ and $\tNPR$ now have the same exit waiting times, the distribution of waiting times for {\R} is given by $\fra=\fpua$, with corresponding survival function $\Sra$.

These two steps lead to the model in \figref{flowmodels}C. It is now possible to recycle our preceding analysis and write down expected counts in these new classes. Survival in the state of being \emph{simultaneously} alive and \emph{recently infected}, is given by $\Sr(t)=\Sa(t)\Sra(t)$, and hence for the progressing populations we obtain
\begin{align}
    \PU(t)&=(1-\Pnp)\int_{-\infty}^{t}I(s)\H(s)\Sr(t-s)\,ds\label{PUeq}\\
    \intertext{and}
    \PO(t)&=(1-\Pnp)\int_{-\infty}^{t}I(s)\H(s)\Sa(t-s)(1-\Sra(t-s))\,ds.
\end{align}
Note the similarity with expressions for $\U(t)$ and $\O(t)$ in the basic model. For the non-progressing populations we obtain
\begin{align}
    \NPR(t)&=\Pnp\int_{-\infty}^{t}I(s)\H(s)\Sr(t-s)\,ds\label{NPReq}
    \intertext{and}
    \NPL(t)&=\Pnp\int_{-\infty}^{t}I(s)\H(s)\Sa(t-s)(1-\Sra(t-s))\,ds.
\end{align}
For convenience we define
\begin{equation}
    \epsilon=\frac{\Pnp}{1-\Pnp},
\end{equation}
and note that
\begin{align}
    \NPR(t)&=\epsilon\PU(t)\label{nplitoPU}
    \intertext{and, more importantly,}
    \NPL(t)&=\epsilon\PO(t).\label{nplitoPO}
\end{align}
These equations express the symmetry between the progressing and
non-progressing sub-populations of \figref{flowmodels}C.
Substituting \eqref{PUeq} and \eqref{NPReq} into \eqref{Req}, we can write
\begin{equation}
    R(t)=\int_{-\infty}^{t}I(s)\H(s)\Sr(t-s)\,ds.
\end{equation}
It is appropriate to use a weighting scheme analogous to the one used in the basic model
\begin{equation}
    W(s,t)=\H(s)\Sr(t-s),
\end{equation}
since $W(s,t)$ is now proportional to the probability that individuals are alive and classified as \emph{recently infected}
at time $t$ if they become infected at time $s$, regardless of whether they are progressors or
non-progressors. Then the weighted incidence, denoted $\Iw$, is given
by
\begin{align}
    \Iw(t)&=\frac{\int_{-\infty}^{t}I(s)\H (s)\Sr(t-s)\,ds}{\int_{-\infty}^{t}\H(s)\Sr(t-s)\,ds}\notag\\
    &=\frac{R(t)}{\int_{-\infty}^{t}\H(s)\Sr(t-s)\,ds}.\label{exacteq}
\end{align}
The populations of \emph{under}-threshold (\U) and
\emph{over}-threshold (\O) individuals are related to the
populations defined in \figref{flowmodels}C by
\begin{equation}
    \U(t)=\PU(t)+\NPR(t)+\NPL(t)
\end{equation}
and
\begin{equation}
    \O(t)=\PO(t).
\end{equation}
Using the above two equations and \eqref{Req} and \eqref{nplitoPO} this means that the population of recent infections is related to
the under-threshold and over-threshold populations by
\begin{equation}
    \R(t)=\U(t)-\epsilon\O(t).
\end{equation}
Performing the same expansion technique as before and assuming a slowly varying healthy population gives the simple expression
\begin{equation}
    \Iw(t)=\frac{\U(t)-\epsilon\O(t)}{\H(t)\,\E{\taur}}.\label{maininfformula}
\end{equation}
This expresses the incidence in terms of the calibration
parameters $\E{\taur}$ and $\epsilon$ (equivalently $\Pnp$), and the state variables $\H$, $\U$ and $\O$.

All that has changed, as a result of allowing non-progressors into
the model, is the shift in the numerator from $\U$ to $\R=\U-
\epsilon\O$. The same bias calculations as before apply
immediately, but there is an increase in statistical sensitivity.
This can be understood by noting that the gross error in $\U$
becomes the dominant part of a fractional error in $\R$ and that
$\R$ is smaller than $\U$.



\section{Statistics and Calibration}\label{calibration}

The population models of the preceding section are expected to be in ever
closer correspondence to a real population as the population size
increases. To model the sampling process of a cross-sectional
survey with a sample size $N$, we rescale the sub-populations of
the continuous time model, at any time $t$, by the total population size
$T=\H+\U+\O$, to obtain the population proportions $\PrH=\H/T$,
$\PrU=\U/T$ and $\PrO=\O/T$. The result of a survey employing the
TRI is the set of three counts $\NH+\NU+\NO = N$. These counts are
trinomially distributed around their means $\bar \NH=\PrH N$,
$\bar \NU=\PrU N$ and $\bar \NO=\PrO N$. These observed counts
turn equation \eqref{maininfformula} into an estimator for the
recently weighted incidence $\Iest$ given by
\begin{equation}
    \Iest=\frac1{\E{\taur}}\frac{\NU-\epsilon \NO}{ \NH}.\label{simpleestimator}
\end{equation}
It should be noted that we do not address issues relating to experiment design and selection bias.

\subsection*{Statistical Fluctuations}

As noted in the preceding section, in a relatively established
epidemic where the smallest class is {\U}, the major source of
fluctuations in $\Iest$ is $\NU$. Crudely speaking then, we can
estimate the reproducibility by blaming all the statistical
uncertainty on the measurement of $\NU$, which has a standard
deviation $\sigma(\NU)=\sqrt{N\PrU(1-\PrU)}\approx\sqrt{N\PrU}$. This leads to the
`back of the envelope' formula for the relative standard deviation
given by
\begin{equation}
    \frac{\sigma(\Iest)}{\Iest}\approx\frac{1}{\PrU-\epsilon\PrO}
    \sqrt{\frac{\PrU}{N}}.\label{trivialdeltaioveri}
\end{equation}
However, the trinomial counts in the estimator \eqref{simpleestimator}
fluctuate and are correlated since they are constrained to add up
to the sample size $N$. In the Appendix we demonstrate how these counts can be modeled by two \emph{independent} draws
($\alpha_1$ and $\alpha_2$) from a standard normal distribution.
We obtain a particular incidence estimator by inserting the
counts, as functions of $\alpha_1$ and $\alpha_2$ into
\eqref{simpleestimator}. Organizing the resulting expression into
a natural expansion in powers of $1/\sqrt{N}$ gives
\begin{equation}
    \Iest(\alpha_1,\alpha_2)=\Iest(0,0)-\frac{1}{\sqrt{N}}
    \left[\alpha_1\frac{\gamma_1(\PrU-\epsilon\PrO)}{\PrH^2(\PrO+\PrU)}
    +\alpha_2\frac{\gamma_2(1+\epsilon)}{\PrH}\right]\frac1{\E{\taur}}
    +O\left(\frac{1}{N}\right),\label{Iwitoalphas}
\end{equation}
where 
\begin{equation}
    \gamma_1=\sqrt{\PrH(1-\PrH)}\qquad\mbox{and}\qquad\gamma_2=\sqrt{\frac{\PrO\PrU}{\PrO+\PrU}}
\end{equation}
The leading term,
\begin{equation}
    \Iest(0,0)=\frac1{\E{\taur}}\frac{\bar \NU-\epsilon\bar \NO}{\bar \NH}
    =\frac1{\E{\taur}}\frac{\PrU-\epsilon\PrO}{\PrH},\label{fracest}
\end{equation}
is just the estimator evaluated at the expectation values of the
counts. The $O(1/N)$ term, which we have omitted, contains a term
proportional to $\alpha_1^2$. This means that there is in
principle a finite sample size bias in the estimator, which is
however suppressed by $O(1/N)$ relative to the dominant term, as
is borne out by numerical calculations in Section~\ref{numerical}.
The retained sub-leading term (of order $O(1/\sqrt{N})$) is
distributed according to a bivariate normal distribution. Thus, to
this order, there is no bias and we can read off the likelihood of observing a value of $\Iest$,
\begin{equation}
    L(\Iest)=f\left(\frac{\Iest-\Iest(0,0)}{\sigma(\Iest)}\right),
\end{equation}
where $f(\cdot)$ is the standard normal density and
\begin{equation}
    \frac{\sigma(\Iest)}{\Iest(0,0)}=\sqrt{\frac{1}{N}\frac{1}{\PrO+\PrU}
    \left(\frac{1}{\PrH}+\frac{\PrO\PrU(1+\epsilon)^2}{(\PrU-\epsilon\PrO)^2}\right)}.
\label{systematicdeltaioveri}
\end{equation}

Comparison with numerical simulation suggests that this
approximate result is essentially indistinguishable from the exact
answer in the regime of realistic values for $N$ and the proportions
($\PrH$, $\PrU$ and $\PrO$), given that in practice one uses the
sample proportions as estimates of the population proportions. \figref{reproduce} plots the relations \eqref{trivialdeltaioveri}
and \eqref{systematicdeltaioveri} for different values of $\Pnp$
and fixed values of $\PrH$, $\PrO$ and $\PrU$. Comparison with
\figref{bias} confirms that the truncation of the expansion of the
healthy population to a constant term (the crucial step in
obtaining the simple incidence relation on which the estimator is
based) leads to a bias that is small compared to realistic
statistical errors. The close correspondence between the two
curves suggests that the simple formula \eqref{trivialdeltaioveri}
should be sufficiently accurate to choose a sample size for an
intended study.

\subsection*{Calibration}

Aside from the sample counts $\NH$, $\NU$ and $\NO$, all other
quantities in relations of the kind derived in the previous
section, such as $\E{\taur}$ in the estimator, should be regarded
as parameters that need to be estimated
independently of a cross-sectional survey used to infer incidence.
Even in the more general case, where the healthy population is
allowed to vary considerably over the time when the weighting
function is non-vanishing, calibration consists only of estimating
$\Pnp$ and expressions of the form  $\E{\taur^{n}}$. We have
already remarked that for the BED assay it will probably not be possible to obtain
reasonable estimates of $\E{\taur^{n}}$ for values of $n$ other
than $1$ and that it appears that only this term is really needed
for practical purposes.

Note that if one wishes merely to estimate \emph{trends} in
incidence, as opposed to absolute incidence values, then it is not
necessary to have an estimate for $\E{\taur}$ at all, since it is
just an overall factor. If the overall scale of incidence
estimates is to be known, considerable effort will need to be
invested in the estimation of $\E{\taur}$. Note that this is the mean time progressing individuals spend under the threshold, with mortality accounted for.


However, surveys conducted at different times will not yield
comparable values of $\Iest\propto(\U-\epsilon\O)/\H$ unless $\epsilon$ (equivalently $\Pnp$) is known with some accuracy, since it appears in one of two terms in the numerator. Consider two surveys which use the same point estimate
\begin{equation}
    \epsilon=\epsilon_0+\epsilon_1,
\end{equation}
where $\epsilon_0$ is the real value and $\epsilon_1$ is the error due to methodological and statistical factors. The first survey obtains values of $\NH^{(1)}$, $\NU^{(1)}$ and $\NO^{(1)}$ and the second obtains values of $\NH^{(2)}$, $\NU^{(2)}$ and $\NO^{(2)}$. This leads to an estimate of the
difference between the two incidences of
\begin{equation}
    \Delta\Iest(\epsilon)=\Iest^{(1)}(\epsilon)-\Iest^{(2)}(\epsilon)
    =\frac{\NU^{(1)}-\epsilon \NO^{(1)}}{\NH^{(1)}\E{\taur}}
    -\frac{\NU^{(2)}-\epsilon \NO^{(2)}}{\NH^{(2)}\E{\taur}}.
\end{equation}
Knowledge of the exact value $\epsilon_0$ leads to
\begin{equation}
    \Delta\Iest(\epsilon_0)=\Iest^{(1)}(\epsilon_0)-\Iest^{(2)}(\epsilon_0)
    =\frac{\NU^{(1)}-\epsilon_0 \NO^{(1)}}{\NH^{(1)}\E{\taur}}
    -\frac{\NU^{(2)}-\epsilon_0 \NO^{(2)}}{\NH^{(2)}\E{\taur}},
\end{equation}
from which we see that the error in $\Delta \Iest$, due to the
error in $\epsilon$, is
\begin{equation}
    \Delta\Iest(\epsilon)-\Delta\Iest(\epsilon_0)
    =\frac{\epsilon_1}{\E{\taur}}\left(\frac{\NO^{(2)}}{\NH^{(2)}}
    -\frac{\NO^{(1)}}{\NH^{(1)}}\right).\label{epserror}
\end{equation}
The direction and magnitude of error depend in detail on many
factors, such as population renewal and long term post-infection
survival. While it is not possible to summarize all the effects
that may be produced by imperfect estimation of $\Pnp$, in
Section~\ref{numerical} we conduct a number of numerical
simulations which demonstrate the kind of bias that may arise.



\section{Numerical Simulations}\label{numerical}

In this section we briefly outline two numerical simulations. The
first serves to test the accuracy of the bias and standard
deviation estimates derived from the truncation of the systematic
expansion of the stochastic estimator $\Iest(\alpha_1, \alpha_2)$. For each of 10,000,000 iterations, two standard normal variables were drawn. Counts for healthy, under-threshold and over-threshold sub-populations, within a sample of size $N=5000$, were generated according to the procedure provided in the Appendix. This algorithm incorporates the assumption that the trinomial distribution of the sample proportions can be approximated as normal, but involves no truncation of the estimator in powers of $1/\sqrt{N}$. The resulting ensemble of point estimates $\Iest$ produced suitably converged
estimates for the mean and standard deviation.

The entire procedure of the preceding paragraph was repeated for
values of $\PrO \in [0.1,0.5]$. For each value of $\PrO$, the
value of $\PrU$ was varied to produce an incidence in the range
$[0.01,0.2]$. The average
fractional discrepancy between the mean incidence and $\Iest(0,0)$
was 0.0003, with values ranging from 0.00001 to 0.0008, confirming
that the intrinsic sampling bias is of order $1/N$. The average
fractional discrepancy between the observed standard deviation of
the incidence and the approximate expression
\eqref{systematicdeltaioveri} was 0.0006, with values ranging from
0.00004 to 0.001, which is also consistent with the truncation at
$O(1/\sqrt{N})$.

A second simulation demonstrates the use of the simple expression \eqref{simpleestimator} for incidence estimation. Arrival times of new infections were generated according to a non-homogeneous Poisson process with intensity given by $\lambda(t)=\H(t)I(t)$ as described in more detail below. Newly infected individuals were initially classified as under the recency threshold of a TRI. A fraction
$1-\Pnp$ progressed to the over-threshold category according to waiting times generated by $\fra$. Weibull functions were used for the waiting time distributions $\fra$ and $\fa$. Unique individuals were drawn from the population at intervals, to produce counts $\NH$, $\NU$ and $\NO$, and hence estimates for incidence.

To demonstrate the incidence estimation process, a $50$ year
population scenario was produced. \figref{distributions} shows the
Weibull distributions used for $\fa$ and $\fra$ (top) and the
corresponding survival curves $\Sa$ and $\Sra$ (bottom). The
Weibull shape and scale parameters for the distributions were
chosen to give approximately realistic values for the mean and
standard deviations for the window period and infected life
expectancy, as detailed in Table~\ref{tableW}. The healthy
population was set to $\H(t) = 100,000 + 5,000t$, with $t$ measured
in years. The incidence was set at $0.01$ (hazard per person per
year) for the first ten years, climbing linearly to 0.1 over the
next ten years, then remaining at this high level for a further
ten years, followed by ten years of linear decline to $0.03$ and
maintained at this level for the last ten years of the simulation.

\begin{table}[!h]
\centering
\begin{tabular}{|l|l|l|l|l|}
    \hline
    & Shape ($k$) & Scale ($l$) & Mean & Standard Dev.\\
    \hline
    Life expectancy ($\fa$) & 4.5 & 8.83 & 8 years & 2 years\\
    Window period ($\fra$) & 8 & 0.58 & 200 days & 30 days\\
    \hline
\end{tabular}
\parbox{12.5cm}{\caption{Weibull parameters for the Monte Carlo
simulation (survival time measured in years).}\label{tableW}}
\end{table}

\figref{montecarlo} shows output from this simulation. The input
incidence parameter is indicated as the dotted \emph{instantaneous
incidence} curve. A sample of $5,000$ individuals was surveyed
every year, and an incidence estimate was produced using the
simple estimator \eqref{simpleestimator} with exact values of
$\E{\taur}$ and $\Pnp$, i.e., assuming perfect calibration. These
point estimates are indicated as \emph{estimated incidence}
values, using `$+$' symbols. The combined effects of the
previously noted time convolution, in $\Iw$, as well as stochastic
departure from means in the simulated population, make the input
incidence parameter an unrealistic target for simulated incidence
measurements. Thus, the solid \emph{weighted incidence} line has
been displayed, which uses full knowledge of all population
members' classification into {\H}, {\U} or {\O}, inserted into
\eqref{exacteq} with full knowledge of the denominator,
(both the non-constant $\H(t)$ and the exact $\Sr$). This is
essentially all that the incidence estimation algorithm can be
asked to reproduce. A \emph{two standard deviation envelope}
around the \emph{weighted incidence} line, calculated from
\eqref{systematicdeltaioveri} using knowledge of the full
population, is shown as two dashed lines.

In Section~\ref{calibration} it was shown that incidence trends
can be extracted without $\E{\taur}$ calibration, while an
estimate for $\Pnp$ is vital. We now explore the extent to which
the accuracy of the estimate of $\Pnp$ affects the ability to
determine a trend in incidence.

Population fractions for {\H}, {\U} and {\O} were
extracted at six times from the population simulation described
above and are shown in Table~\ref{tablePr}. Four instances of
incidence trend estimation were simulated by selecting the time
intervals $(15,20)$, $(20,30)$, $(30,35)$ and $(40,50)$. We
considered the trends that would be observed if incidence were
measured at the beginning and end of each of these intervals using
\eqref{fracest}. In order to focus on the bias introduced by
imperfect estimation of $\Pnp$, rather than sample size effects,
we assumed perfect knowledge of $\PrH$, $\PrU$ and $\PrO$. For each
of these intervals, we calculated an incidence estimate at the
beginning and end, as a function of the estimated value of $\Pnp$
(the true value being $0.05$), assuming $\E{\taur}$ is known
exactly. We also calculated the estimated fractional change in
incidence over each interval. Note that the fractional change does
not depend on $\E{\taur}$. The results are shown in
\figref{calbias}, where the four intervals $(15,20)$, $(20,30)$,
$(30,35)$ and $(40,50)$ are referred to as scenarios A, B, C and
D, respectively.

In each case, the effect of the error in the estimation of $\Pnp$ is quite different, as can be understood by considering how \eqref{epserror} is impacted by the system history. Note that case B and case D both simulate intervals over which incidence is approximately constant, but the impact (on the estimated incidence change) of incorrect estimation of $\Pnp$ does not even agree in sign. At a time of $40$ years, the incidence estimate becomes negative when the $\Pnp$ estimate is higher than $0.0882$. This breakdown of the model results in the divergence of the fractional change in estimated incidence over the interval in scenario D. In short, incorrect estimates of $\Pnp$ can lead to the fundamental breakdown of the inference scheme. This makes sense, as $\Pnp$ impacts the long term accumulation of individuals in the $\PrO$ category.

\begin{table}[!h]
\begin{center}
\begin{tabular}{|c|c|c|c|c|c|c|}
    \cline{2-7}
    \multicolumn{1}{c|}{} & \multicolumn{6}{|c|}{Time (years)}\\
    \multicolumn{1}{c|}{} &  \multicolumn{1}{c}{15} & \multicolumn{1}{c}{20} &
    \multicolumn{1}{c}{30} & \multicolumn{1}{c}{35} & \multicolumn{1}{c}{40} & \multicolumn{1}{c|}{50}\\
    \hline
    $\PrH$ & 0.849 & 0.687 & 0.576 & 0.602 & 0.694 & 0.814\\
    $\PrU$ & 0.030 & 0.050 & 0.051 & 0.041 & 0.027 & 0.022\\
    $\PrO$ & 0.121 & 0.263 & 0.373 & 0.357 & 0.279 & 0.164\\
    \hline
\end{tabular}
\parbox{12.5cm}{\caption{Population fractions in {\H}, {\U} and {\O} within a 50 year epidemic scenario.}\label{tablePr}}
\end{center}
\end{table}



\section{Discussion and Conclusion}

We have presented a detailed analysis of relations between recent
incidence in a population and counts of `recently infected'
individuals. These are in principle complex convolutions involving
the epidemiological history as well as all information about the
distribution of waiting times in the recently infected category.
When the healthy population undergoes realistically modest
variation on the time scale of the definition of recency implied
by the TRI, we obtain simplified forms which incur very little
error. The simplified relations form the basis of estimators which
are shown to have considerably more variance than bias under
realistic demographic and epidemiological assumptions.

Noting that the assumptions of our model are the least restrictive of any BED based HIV incidence estimation method of which we are aware, we now consider its limitations. We have only modeled one direction of progression of individuals from an experimentally defined state of `recent' infection to `non-recent' infection.
The reverse apparently occurs for BED optical density in some
terminal stage AIDS patients. This process constitutes a
substantial complication, and further work is required to
investigate how it may be incorporated into an analytical model of
the kind developed here. It may be worth exploring previous
suggestions \cite{bed:2006:1} to use additional information, from
questionnaires or other assays, to remove end-stage patients from
the observed recent count. We have also not considered the
possibility that calibration parameters are functions of time, for
example as a result of substantial vaccine uptake in a population.
Even more subtle is a point raised under calibration, namely that
imperfect estimation of the non-progressing fraction of the
population leads to a complex bias in incidence estimates. This
bias is dependent on factors not even present in the incidence
estimator, such as long term survival post infection.

A key observation is that, for the purposes of estimating
incidence from a TRI, there is no fundamental obstacle posed by
having a \emph{known} fraction of individuals fail to progress
over the recency threshold, as long as their distribution of
survival times from infection is the same as in the progressing
population. In fact, an \emph{accurate} estimate of this
non-progressing fraction alone, is sufficient (and necessary) to
infer \emph{trends} in incidence. This fraction could possibly be
estimated for the BED assay from historical records, since there
are many viable samples in storage with supporting clinical
information indicating long-infected status. However, as
demonstrated in the calculations of Section~\ref{numerical}, a
suitably large error in the estimate of $\Pnp$ can render TRI
based incidence estimates meaningless. A calibration of the mean finite waiting time is required in order to estimate \emph{absolute values} of incidence. Prospective follow-up is probably the only practical way to estimate this parameter.


In contrast to our model, which has only two calibration parameters, the model of McDougal et al.~\cite{bed:2006:1} appears to have three (sensitivity, short-term specificity and long-term specificity). In a separate short note \cite{McWalterWelte2008B} we demonstrate that, under their own assumptions, these parameters can be reduced to ours. This has two advantages---our parameters are easier to calibrate and assuming independence of their parameters would lead to incorrect error estimates.

Besides the explicit assumptions noted, the analysis presented
here is quite general. Tests for recent infection continue to be
of interest, and new assays are likely to be developed both for
HIV infection and other important diseases. In summary, we have
presented a simple incidence estimator and a detailed analysis of its likelihood function, which can
inform design of appropriate calibration studies and
cross-sectional incidence estimation surveys, and can also form
the basis of systematic inference algorithms for processing the
data obtained from such surveys.


%% file: appendix.tex
\newcommand{\PropH}{P_{{\scriptscriptstyle \Hsub}}}
\newcommand{\PropU}{P_{{\scriptscriptstyle \Usub}}}
\newcommand{\PropO}{P_{{\scriptscriptstyle \Osub}}}
\newcommand{\PropX}{P_{{\scriptscriptstyle \textrm{X}}}}

\section*{Appendix}

Given a sample of $N$ subjects tested using the TRI, we derive a systematic expansion, in powers of $O(1/\sqrt{N})$, for the estimator
\begin{equation}
    \Iest=\frac1{\E{\taur}}\frac{\NU-\epsilon \NO}{\NH} =\frac1{\E{\taur}}\frac{\PropU-\epsilon \PropO}{\PropH},
\end{equation}
where the counts $\NX=\PropX N$ fluctuate trinomially around their means $\bar{\NX}=\PrX N$, with standard deviations $\sigma_{\scriptscriptstyle\textrm{X}}=\sqrt{N\PrX(1-\PrX)}$, or, alternatively, the realized sample proportions $\PropX$ fluctuate multinomially around their means $\PrX$, with standard deviations $\sigma_{\PropX}=\sqrt{\PrX(1-\PrX)/N}$. In order to account for fluctuations in the sample counts, as well as their correlations, we express the three counts as the result of two independent random draws. We assume the counts are sufficiently large so that binomial distributions can be approximated by normal distributions---which will be the case if the survey is to have any reasonable accuracy.
\begin{itemize}
\item Draw a number $\alpha_1$ from a standard normal distribution
and set
\begin{equation}
    \NH=\bar \NH+\sigma_1\alpha_1,
\end{equation}
where
\begin{equation}
    \sigma_1=\sigma_{\scriptscriptstyle\textrm{S}}=\sqrt{N\PrH(1-\PrH)}=\sqrt{N}\gamma_1.
\end{equation}
Defining $\gamma_1$ allows us to keep track of powers of
$1/\sqrt{N}$.

\item Draw a number $\alpha_2$ from a standard normal distribution
and set
\begin{align}
    \NU&=\PrU^\prime(N-\NH)+\sigma_2\alpha_2\notag\\
    &=\PrU^\prime N(1-\PrH)-\PrU^\prime\sigma_1\alpha_1+\sigma_2\alpha_2\notag\\
    &=\PrU N - \PrU^\prime \gamma_1 \alpha_1 \sqrt{N}  +\sigma_2\alpha_2\\
\intertext{and}
    \NO&=\PrO^\prime(N-\NH)-\sigma_2\alpha_2\notag\\
    &=\PrO^\prime N(1-\PrH)-\PrO^\prime\sigma_1\alpha_1-\sigma_2\alpha_2\notag\\
    &=\PrO N - \PrO^\prime \gamma_1 \alpha_1 \sqrt{N}  -\sigma_2\alpha_2
\end{align}
where
\begin{equation}
\PrU^\prime=\frac{\PrU}{\PrU+\PrO}\qquad\mbox{and}\qquad\PrO^\prime=1-\PrU^\prime
\end{equation}
are the probabilities of an individual being in {\U} and {\O}
given that they are not in {\H}, and
\begin{equation}
    \sigma_2=\sigma_{{\scriptscriptstyle\textrm{B}}|\alpha_1}=\sqrt{(N-\NH)\PrU^\prime\PrO^\prime}
    =\sqrt{N}\gamma_2\left(1+O\left(\tfrac{1}{\sqrt{N}}\right)\right)
\end{equation}
is the standard deviation of $\NU$ (and $\NO$) subject to
$\alpha_1$ having been determined, so that
\begin{equation}
    \gamma_2=\sqrt{\frac{\PrO\PrU}{\PrO+\PrU}}.
\end{equation}
\end{itemize}
The sample proportions can now be written as:
\begin{align}
    \PropH &=\PrH + \tfrac{1}{\sqrt{N}} \alpha_1 \gamma_1 \nonumber  \\
    \PropU &=\PrU-\tfrac{1}{\sqrt{N}}\big(\alpha_1\PrU^\prime \gamma_1 -\alpha_2 \gamma_2 \big) + O\left(\tfrac{1}{N}\right)\nonumber \\
    \PropO &=\PrO-\tfrac{1}{\sqrt{N}}\big(\alpha_1\PrO^\prime \gamma_1 +\alpha_2 \gamma_2 \big) + O\left(\tfrac{1}{N}\right)
\end{align}
Note that this procedure works independently of the order in which
the values of $\NH$, $\NU$ and $\NO$ are assigned from the random
$\alpha$'s. We can explicitly insert this decomposition into the
expression for $\Iest$ to obtain an expression for the estimator
in terms of two independent draws from a standard normal
distribution. Keeping only the $O(1)$ and $O(1/\sqrt{N})$ terms
explicit leads to
\begin{equation}
    \Iest(\alpha_1,\alpha_2)=\Iest(0,0)-\frac{1}{\sqrt{N}}
    \left[\alpha_1\frac{\gamma_1(\PrU-\epsilon\PrO)}{\PrH^2(\PrO+\PrU)}
    +\alpha_2\frac{\gamma_2(1+\epsilon)}{\PrH}\right]\frac{1}{\E{\taur}}
    +O\left(\frac{1}{N}\right),
\end{equation}
where
\begin{equation}
    \Iest(0,0)=\frac1{\E{\taur}}\frac{\PrU-\epsilon\PrO}{\PrH}.
\end{equation}
Two key observations immediately follow:
\begin{itemize}
\item The proposed $\Iest$ is a consistent estimator in the sense
that terms with bias are at least $O(1/N)$ suppressed relative to
the dominant term.

\item To order $O(1/\sqrt{N})$, $\Iest$ is Gaussian with relative
variance
\begin{equation}
    \left(\frac{\sigma(\Iest)}{\Iest(0,0)}\right)^2=\frac{1}{N}
    \frac{1}{\PrO+\PrU}\left(\frac{1}{\PrH}+\frac{\PrO\PrU
    (1+\epsilon)^2}{(\PrU-\epsilon\PrO)^2}\right).
\end{equation}
\end{itemize}

%% file: figures.tex
\newpage

\begin{figure}[h!]
\centering
\includegraphics[width=16cm]{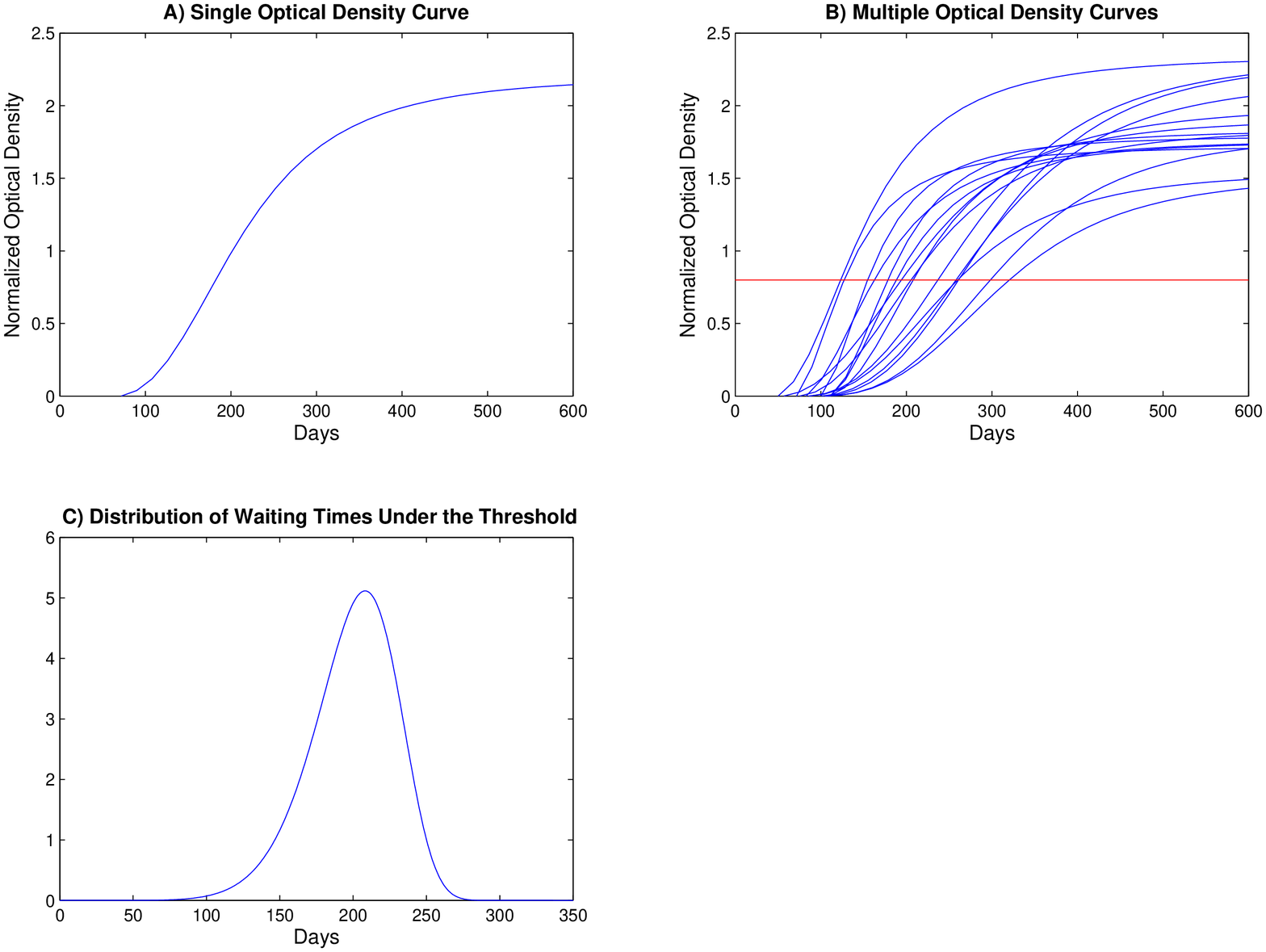}
\caption{} \label{ODcurvedist}
\end{figure}
\vspace{-.65cm}
\begin{enumerate}[A)]
    \item Hypothetical individual BED normalized optical density (ODn) as a function of time since infection.
    \item Hypothetical collection of individual ODn progression plots
        with inter-individual variability and an arbitrary ODn
        threshold which can be used to define \emph{recent
        infection}.
    \item Hypothetical distribution of times spent under the ODn threshold
        after infection, which approximately captures the mean and
        variance of the sample of individual progression curves in B).
\end{enumerate}

$ $\newpage

\input{Flow.TpX}

\vspace{-.65cm}
\begin{enumerate}[A)]
    \item The basic epidemiological/TRI progression model. Members of the
        healthy population {\H} are subject to a per unit time hazard
        (incidence) $I$ of infection. After infection, individuals enter
        the under-threshold population {\U}. Here they spend a time
        distributed according to $\fua$, after
        which they spend the remainder of their lifetime, if any, in the over-threshold population {\O}.
    \item The basic model modified to accommodate non-progressors on the
        TRI. Now, upon infection, a proportion $\Pnp$ of individuals
        remain forever under the threshold of the TRI, i.e., they enter the
        $\tNP$ category. The remaining proportion, $1-\Pnp$, the progressors,
        are assigned a waiting time from $\fpua$, and spend this time in the `progressing, under-threshold' population $\tPU$. Those that survive long enough
        spend the remainder of their lifetime in the `progressing, over-threshold' population $\tPO$
    \item Modified model with separation of non-progressors into `recent'
        and `long' infected categories. This model contains the same epidemiology and
        biology as the model in B), with the introduction of a bookkeeping
        device which facilitates the definition of a calibratable category
        of `recently infected' individuals. The non-progressors are
        assigned waiting times drawn from the distribution observed in the
        progressing population, and spend this waiting time in the `non-progressing recently infected' ($\tNPR$) category, before moving to the `non-progressing long infected' ($\tNPL$) category.
\end{enumerate}

\newpage
\begin{figure}[h!]
\centering
\includegraphics[width=10cm]{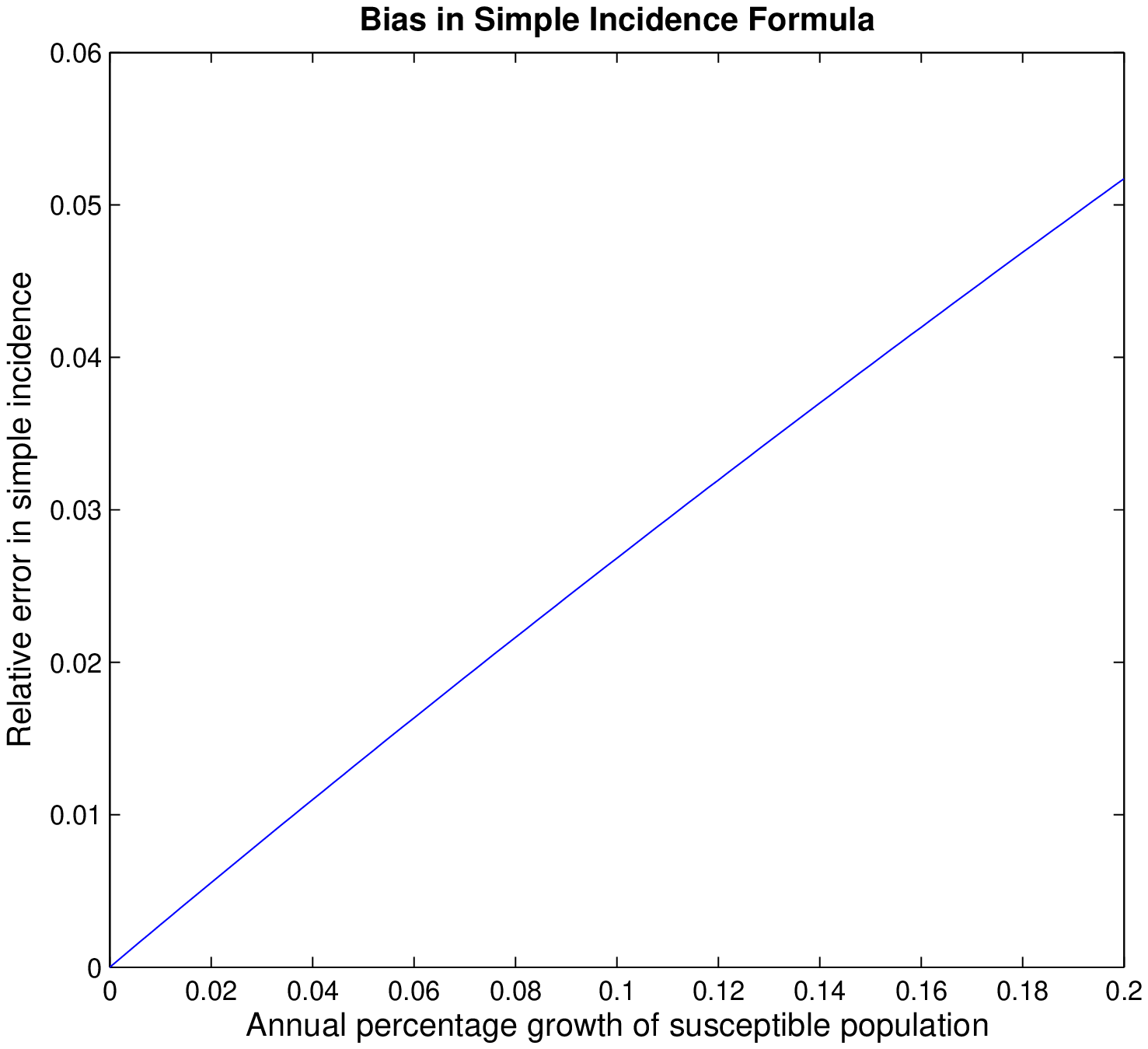}
\caption{Fractional error in the simple incidence relation
\eqref{Isibsimple} versus the full relation \eqref{isibdef}, as a
function of growth rate of the healthy population, quoted as a
percentage annual growth. The scenario is defined by: $S(t) = S(0)
e^{\alpha t}$ and $\fu$ is a Weibull distribution with scale
parameter $l=0.57$ and shape parameter $k=8$. The parameter $\alpha$ is varied
to produce deviation from a constant healthy population ($\alpha=0$) in
which limit equation \eqref{Isibsimple} is exact.}\label{bias}
\end{figure}

\newpage
\begin{figure}[h!]
\centering
\includegraphics[width=10cm]{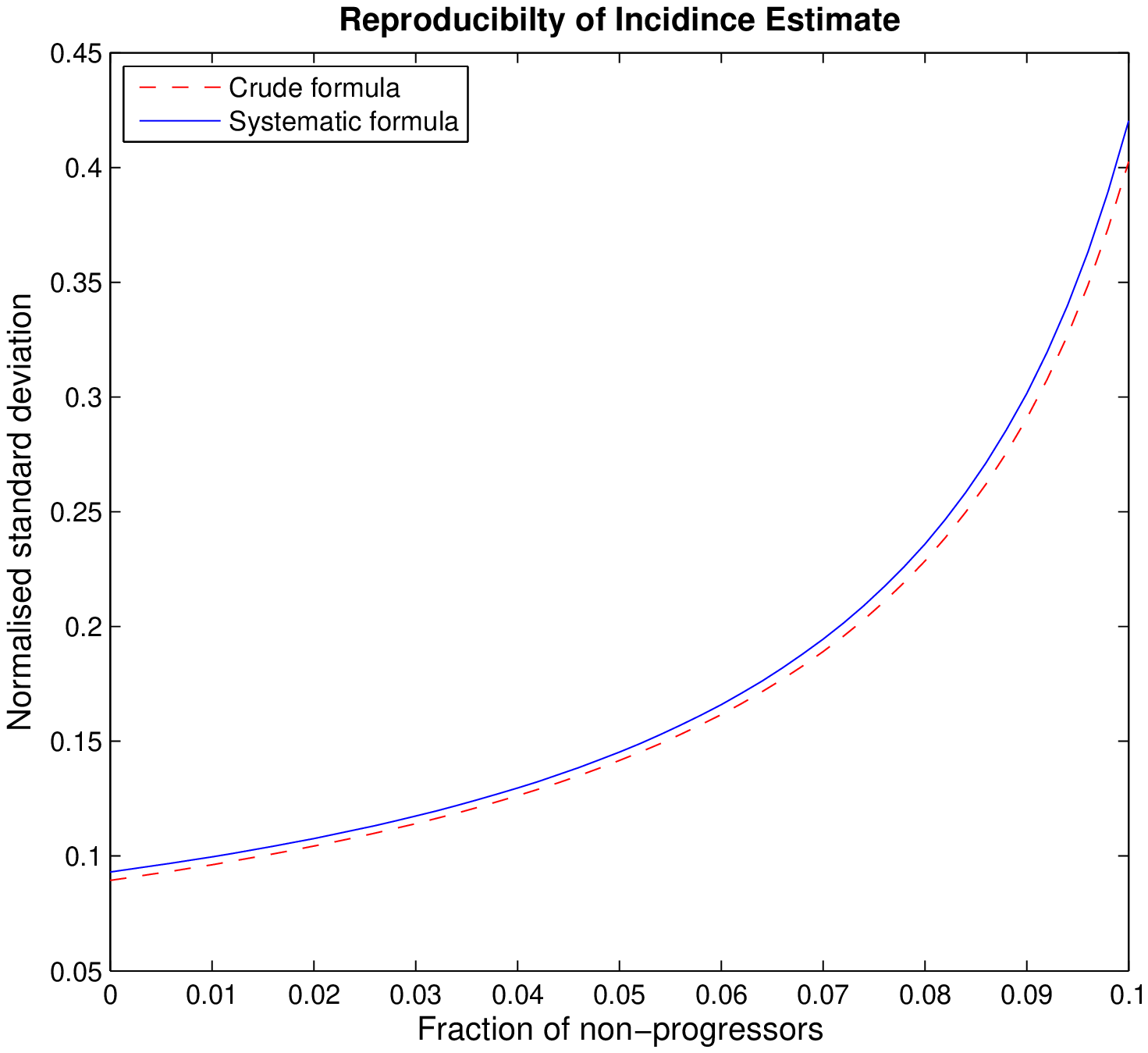}
\caption{Plots of the crude formula \eqref{trivialdeltaioveri}
(dashed line) and the systematic formula
\eqref{systematicdeltaioveri} truncated after $O(1/\sqrt{N})$
(solid line) for the standard deviation of incidence estimates,
expressed as a fraction of the leading term $\Iest(0,0)$, and
plotted as a function of the non-progressing fraction $\Pnp$, for
fixed values $\PrH=0.60$, $\PrU=0.05$ and
$\PrO=0.35$.}\label{reproduce}
\end{figure}

$ $\newpage
\begin{figure}[h!]
\centering
\includegraphics[width=16cm]{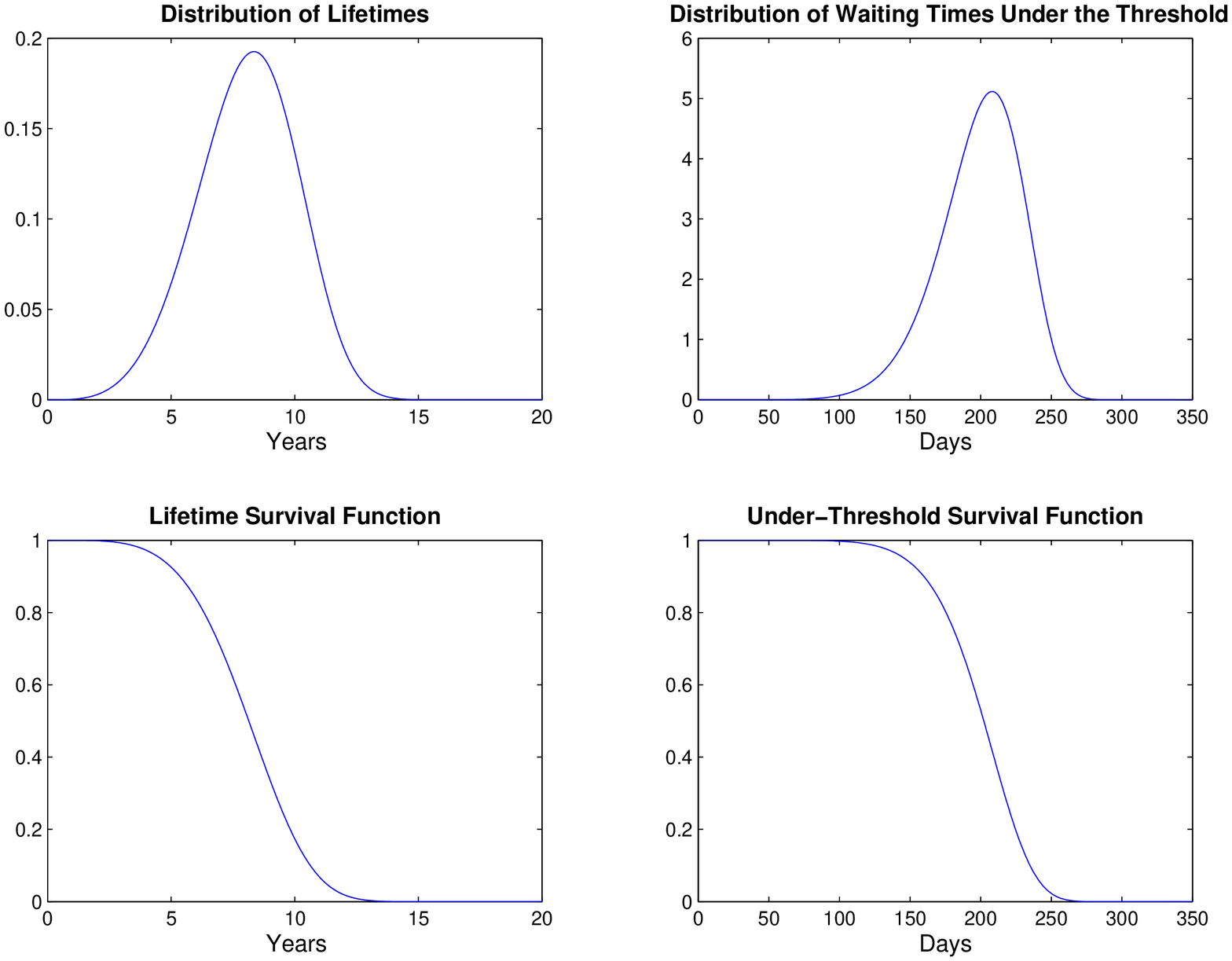}
\caption{Graphical representation of the densities $\fa$ and
$\fra$ for survival times and times under threshold in the progressing population (top) and the
corresponding survival probabilities $\Sa$ and $\Sra$
(bottom). Parameters for these curves are detailed in Table
1.}\label{distributions}
\end{figure}

$ $\newpage
\begin{figure}[h!]
\centering
\includegraphics[width=16cm]{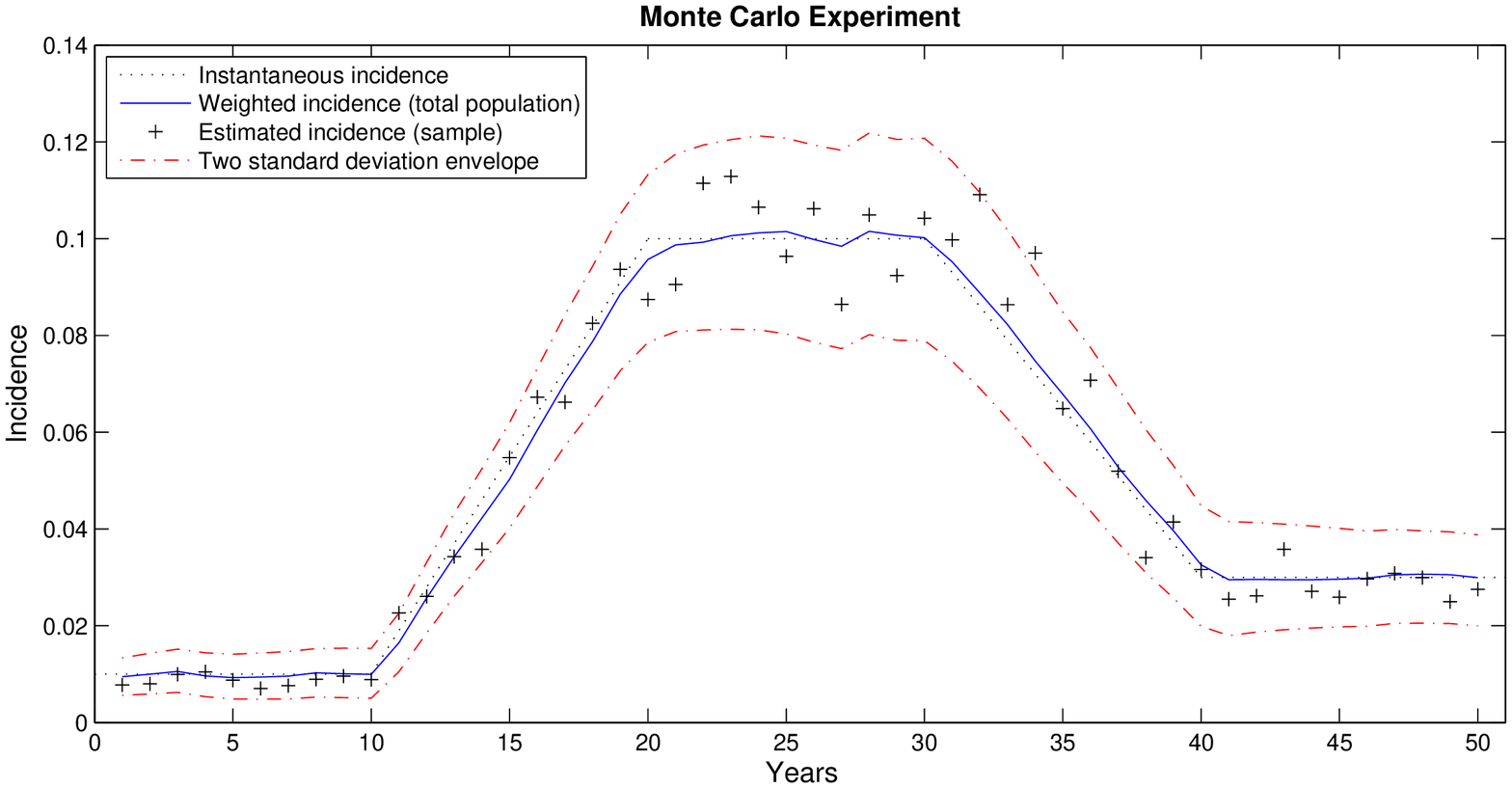}
\caption{Full stochastic simulation of population with epidemic,
individual survival times, and annual sampling of $5,000$
individuals. The healthy population was set to
$\H(t)=100,000+5,000t$ with time measured in years. The
\emph{instantaneous incidence} parameter is the dotted curve. The
target of the estimates is the \emph{weighted incidence} (solid
line), which was calculated explicitly as per
\eqref{exacteq} from all the known inputs. This is flanked
by a \emph{two standard deviation envelope} (dashed lines).
Simulated \emph{estimated incidence} ($+$ symbols) were obtained
by using sample counts in the simple estimator
\eqref{simpleestimator}. The calibration parameters $\E{\taur}$
and $\Pnp$ were assumed to be known exactly.} \label{montecarlo}
\end{figure}

$ $\newpage

\begin{figure}[h!]
\centering
\includegraphics[width=15cm]{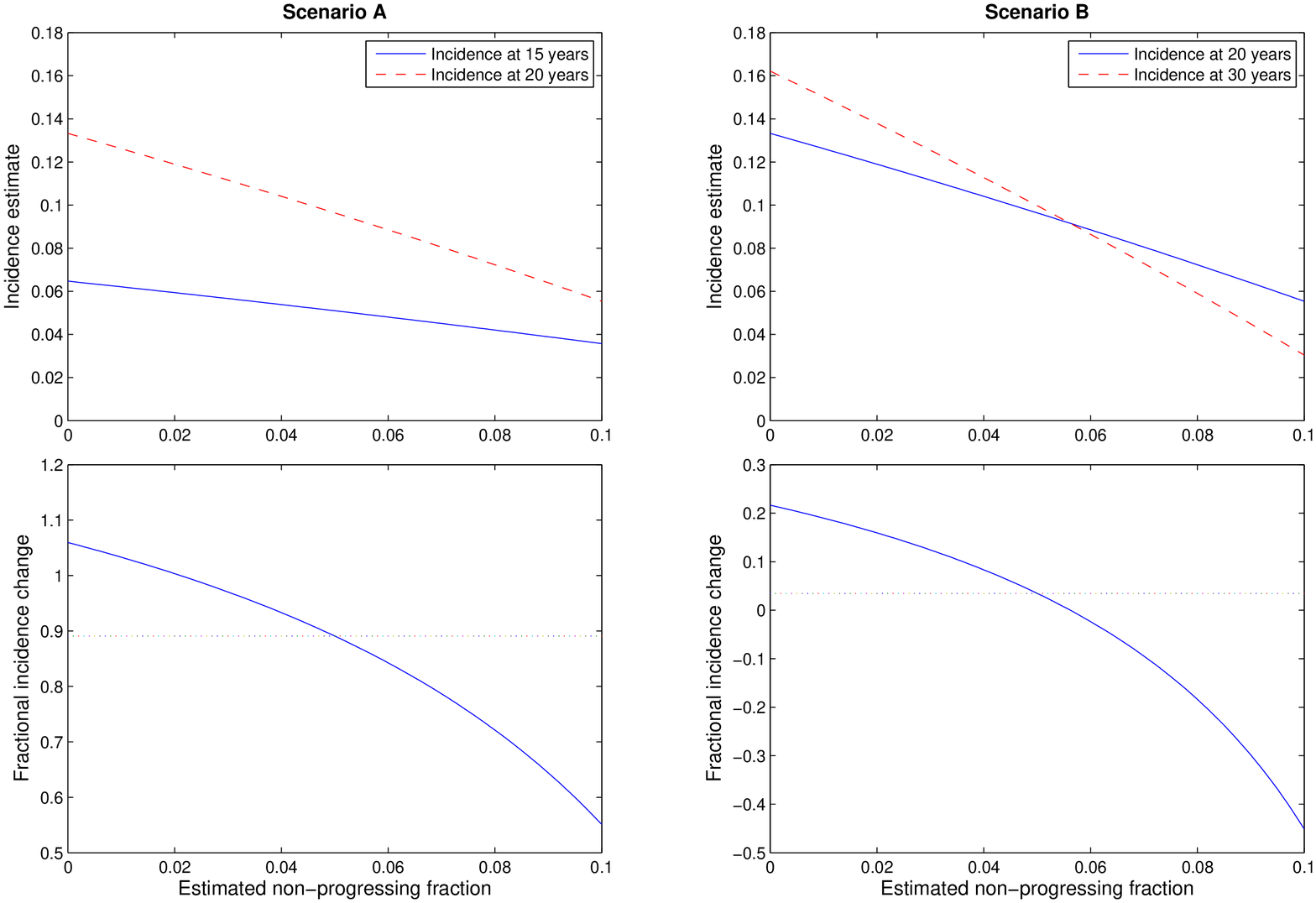}

\vspace{2cm}

\includegraphics[width=15cm]{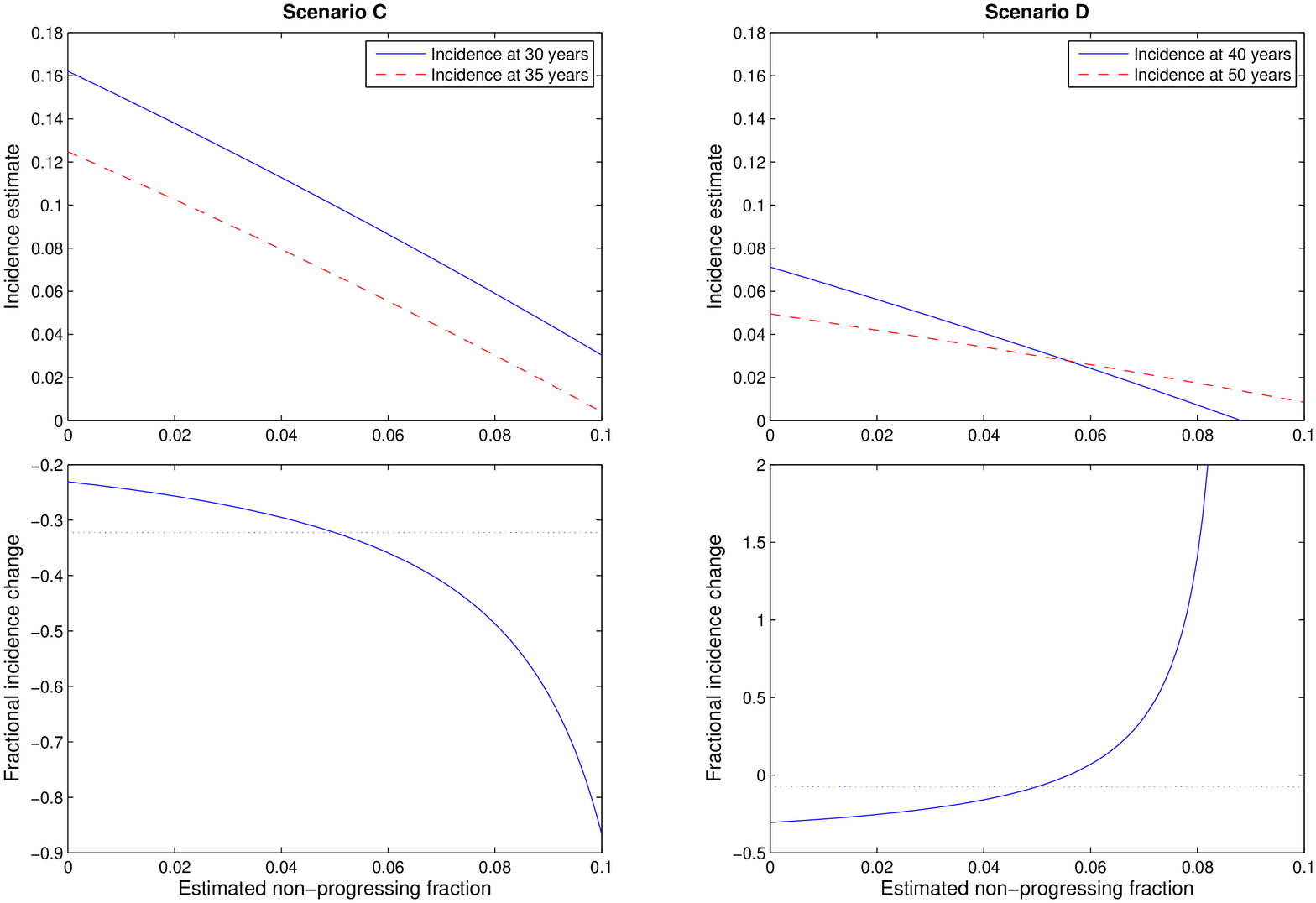}
\caption{Absolute incidence estimates and estimated fractional
incidence changes for four pairs of successive times from
Table~\ref{tablePr}.}\label{calbias}
\end{figure}